\documentclass[11pt]{article}
\usepackage[dvipdfmx]{graphicx}
\usepackage{amssymb,amsmath,bbm,bm,overpic}

\numberwithin{equation}{section}

\begin{document}
\thispagestyle{empty}
\vskip1cm
\begin{center}
{\Large \bf M2-branes and AdS/CFT: A Review}

\vskip1.5cm
Kazuo Hosomichi\footnote{hosomiti@nda.ac.jp}

\bigskip
{\it
Department of Applied Physics, National Defense Academy,\\
1-10-20 Hashirimizu, Yokosuka-city, Kanagawa 239-8686 Japan
}

\end{center}

\vskip1cm
\begin{abstract}
We briefly review some of the important developments in the last decade in the theory of multiple M2-branes and $\text{AdS}_4/\text{CFT}_3$ correspondence. Taking the examples of the superconformal index, free energy on $S^3$ and entropy of charged black holes, we illustrate how the large $N$ limit was studied and the correspondence was checked.
\end{abstract}

\newpage

\setcounter{page}{1}

\section{Introduction}

In 1998, one day when I was a Ph.D. student, Prof.~Eguchi came to me in the tea room and asked if I already read the paper by Maldacena about a new duality which is now known as AdS/CFT correspondence \cite{Maldacena:1997re}. It was when another important paper by Witten \cite{Witten:1998qj} appeared. I had just finished my first paper \cite{Hosomichi:1997if} on emissions from D1-D5 black holes, which was a project suggest by him. Though he did not mean anything special by that small conversation, it remained in my memory because I felt like being treated as an independent researcher for the first time.

In 2009, I started working with Prof. Eguchi again at Yukawa Institute, where he was the Director at that time. We worked together, sometimes jointly with cosmologists, on organizing and running a series of conferences. That was a work requiring a different level of dedication to physics. I was influenced a lot from the eagerness with which he kept these activities running for many years, and also from the way he cared about the purpose and the real outcome for each of those events.

~

In 2019 we had a conference in Kyoto in memory of Prof.~Eguchi. In this article, partly based on the talk given there, I will briefly review some of the important developments in the last decade in the theory of multiple M2-branes and $\text{AdS}_4/\text{CFT}_3$ correspondence. I will illustrate how the large $N$ limit was studied and the correspondence was checked by taking the superconformal index, free energy on $S^3$ and the entropy of charged black holes as examples.

~

Most of the discussions are restricted to the ABJM model \cite{Aharony:2008ug} for $N$ M2-branes probing the orbifold $\mathbb C^4/\mathbb Z_k$. In 3D ${\cal N}=2$ convention, it is a $U(N)_k\times U(N)_{-k}$ Chern-Simons theory with chiral multiplets $A_1,A_2$ in the bifundamental and $B_3,B_4$ in the anti-bifundamental representations and a superpotential
\begin{equation}
 W = -\frac{2\pi}k\text{tr}\left[A_aB_bA_cB_d\right]\epsilon^{ac}\epsilon^{bd}.
\end{equation}
The gauge field, scalar and the auxiliary field in the two $U(N)$ vectormultiplets will be denoted as $(A_\mu,\sigma, D)$ and $(\tilde A_\mu,\tilde\sigma,\tilde D)$, respectively. The model should be dual to the quantum supergravity on $\text{AdS}_4\times S^7/\mathbb Z_k$. The $\text{AdS}_4$ and $S^7/\mathbb Z_k$ have radii $L$ and $2L$, which are related to $N$ and the 11D Newton constant $G_{(11)}$ via
\begin{equation}
 (2\pi\ell_\text{P})^6N=384L^6\cdot\text{vol}(S^7/\mathbb Z_k),\qquad
 16\pi G_{(11)} = \frac{(2\pi\ell_\text{P})^9}{2\pi}.
\end{equation}

\section{Superconformal index}

An important problem in AdS/CFT is to understand the spectrum of states of both sides. Although complete understanding is difficult, precise results can be obtained in supersymmetric theories by restricting the attention to subsectors of states preserving SUSY. One can argue that the index encoding the information of such states is independent of couplings which can vary continuously. Following earlier developments \cite{Kinney:2005ej,Bhattacharya:2008zy} and results \cite{Bhattacharya:2008bja}, an exact formula for the superconformal index was derived for ABJM model in \cite{Kim:2009wb}. The results were shown to agree perfectly with the index over supergravitons in $\mathrm{AdS}_4\times S^7/\mathbb Z_k$ in the large $N$ limit.

\paragraph{Definition.}

The 3D ${\cal N}=6$ superconformal symmetry of the ABJM model has conformal symmetry $SO(2,3)$ and R-symmetry $SO(6)$ as bosonic subgroup. Let us denote by $\epsilon$ and $j_3$ the Cartan generators for $SO(2)\times SO(3)\subset SO(2,3)$, and $h_1,h_2,h_3$ for $SO(6)$. Then one can find nilpotent supercharges $Q$ and $S$ satisfying
\begin{equation}
\left\{Q,S\right\} = \epsilon-h_3-j_3\,,
\end{equation}
and both commuting with $h_1,h_2$ and $\epsilon+j_3$. The superconformal index is defined by the trace
\begin{align}
 I(x,y_1,y_2)&\equiv\text{Tr}\left[(-1)^Fe^{-\beta'\{Q,S\}-\beta(\epsilon+j_3)-\gamma_1h_1-\gamma_2h_2}\right]
\nonumber \\
&=\text{Tr}\left[(-1)^Fe^{-(\beta+\beta')\epsilon-(\beta-\beta')j_3+\beta'h_3-\gamma_1h_1-\gamma_2h_2}\right]
\end{align}
over the Hilbert space of radial quantization. Here $x\equiv e^{-\beta}, y_1\equiv e^{-\gamma_1},y_2\equiv e^{-\gamma_2}$. Note that it is independent of $\beta'$ since it only receives contributions from the states annihilated by $Q$ and $S$.

The index can be computed as a path integral of the theory on $S^1\times S^2$ with the $S^1$ parametrized by Euclidean time $\tau\sim\tau+\beta+\beta'$. The presence of $j_3$ and $h_i$'s in the trace translates into twists in the periodicity of the fields. If one prefers to work with periodic fields, one can take account of them by turning on background $SO(6)$ gauge fields and off-diagonal metric components.

\paragraph{Computation.}

The index can be evaluated with the help of SUSY localization. The path integral
\[
 I \equiv \int {\cal D}(\text{fields}) e^{-(\text{action})}
\]
is supersymmetric; namely there is a supercharge $\bf Q$ under which the measure ${\cal D}(\text{fields})$ and the action are both invariant. As such, $I$ is invariant under modification of the action by terms of the form $\frac1{g^2}{\bf Q}\Psi$, with $\Psi$ fermionic and ${\bf Q}^2\Psi=0$. By choosing $\Psi$ suitably and taking the weak coupling limit $g^2\to0$, one can show the path integral is given exactly by the sum over contributions of saddle points, and that the contribution of each saddle point can be evaluated using Gaussian approximation.

For the ABJM superconformal index, the saddle points are labeled by integers $n_i,\tilde n_i$ and periodic variables $\alpha_i,\tilde\alpha_i$ $(i=1,\cdots,N)$. They appear in the value of flux and temporal holonomy as follows,
\begin{alignat}{2}
\sigma &= \int_{S^2}\frac F{2\pi} = \text{diag}(n_1,\cdots,n_N),
\qquad&
\text P\exp i\int_{S^1}A &=\text{diag}(e^{i\alpha_1},\cdots,e^{i\alpha_N}),
\nonumber \\
\tilde\sigma &= \int_{S^2}\frac{\tilde F}{2\pi} = \text{diag}(\tilde n_1,\cdots,\tilde n_N),
\qquad&
\text P\exp i\int_{S^1}\tilde A &=\text{diag}(e^{i\tilde\alpha_1},\cdots,e^{i\tilde\alpha_N}).
\end{alignat}

The value of the action at this saddle point is
\begin{equation}
 e^{-S} = e^{ik\sum_i(n_i\alpha_i-\tilde n_i\tilde\alpha_i)}.
\end{equation}
This is multiplied by two ``determinants'' to make up the contribution of a given saddle point. Note, as it turns out, that both determinants are invariant under simultaneous shift of the 2$N$ variables $\alpha_i,\tilde\alpha_i$ by the same amount. The integration over $\alpha_i,\tilde\alpha_i$ along this direction thus gives rise to a constraint $\sum_in_i=\sum_i\tilde n_i$.

One of the determinants is the Faddeev-Popov determinant. The flux $(n_i,\tilde n_i)$ generically breaks the gauge group $U(N)\times U(N)$ to a subgroup $\prod_iU(N_i)\times\prod_iU(\tilde N_i)$ with $\sum_iN_i=\sum_i\tilde N_i=N$. The saddle point condition requires the holonomy to take values in this subgroup. Gauge-fixing the holonomy to be also diagonal gives rise to a factor $\frac1{\text{Sym}}\cdot\Delta_\text{FP}$, where
\begin{align}
\text{Sym}&=\prod_iN_i!\prod_i\tilde N_i!,
 \nonumber \\
\Delta_\text{FP} &=
\prod_{i<j(n_i=n_j)}\Big[2\sin\left(\tfrac{\alpha_i-\alpha_j}2\right)\Big]^2
\prod_{i<j(\tilde n_i=\tilde n_j)}\Big[2\sin\left(\tfrac{\tilde\alpha_i-\tilde\alpha_j}2\right)\Big]^2\,.
\label{SD1}
\end{align}

The other is the one-loop determinant arising from Gaussian integration over fluctuation of fields. It can be computed by KK reducing the free theory of fluctuations along $S^2$. The resulting system can be regarded as a bunch of simple bosonic and fermionic harmonic oscillators with periodic Euclidean time. The determinant is its partition function
\begin{equation}
 \Delta_\text{1-loop} =
 \left(\prod_{a:\text{Fermi}}2\sinh\frac{\beta\omega_a}2\right)\Bigg/
 \left(\prod_{a:\text{Bose}}2\sinh\frac{\beta\omega_a}2\right),
\end{equation}
where we used an abbreviation
\begin{equation}
 \beta\omega_a\equiv \beta(\epsilon+j_3)+\beta'(\epsilon-h_3-j_3)+\gamma_1h_1+\gamma_2h_2+(\text{gauge})
\end{equation}
for the $a$-th bosonic or fermionic oscillator. The term (gauge) represents the gauge charge: for example it is $\alpha_i-\tilde\alpha_j$ if the oscillator originates from the $(i,j)$-component of a bi-fundamental field. For later use, we rewrite it into a plethystic exponential
\begin{equation}
 \Delta_\text{1-loop} = \exp\bigg[-\beta\epsilon_0+\sum_{n\ge1}\frac1nf(x^n,y_1^n,y_2^n,e^{in\alpha_i},e^{in\tilde\alpha_i})\bigg],
\end{equation}
where the Casimir energy $\epsilon_0$ and the letter index $f$ are defined by
\begin{equation}
 \beta\epsilon_0 \equiv \sum_{\text{B}\,-\,\text{F}}\frac{\beta\omega_a}2,
 \qquad
 f(x,y_1,y_2,e^{i\alpha_i},e^{i\tilde\alpha_i})\equiv \sum_{\text{B}\,-\,\text{F}}e^{-\beta\omega_a}.
\end{equation}
The KK reduction is performed using monopole harmonics (spherical harmonics for charged fields in the flux background). The quantities $\epsilon_0$ and $f$ therefore depend on the flux $n_i$ and $\tilde n_i$ as well, though not indicated explicitly.

The contribution to the index from saddle points with flux $(n_i,\tilde n_i)$ is thus given by
\begin{equation}
 I\Big|_{(n_i,\tilde n_i)}~=~
\frac1{\text{Sym}}\int\prod_{i=1}^N
\frac{\textrm d\alpha_i\textrm d\tilde\alpha_i}{(2\pi)^2}\cdot
\exp\left[ik\sum_{i=1}^N(n_i\alpha_i-\tilde n_i\tilde \alpha_i)
-\beta\epsilon_0+\sum_{n\ge1}\frac1nf(\cdot^n) \right],
\label{In}
\end{equation}
where the quantity $\text{Sym}$ is defined in (\ref{SD1}), and the letter index $f$ takes account of both the Faddeev-Popov and one-loop determinants.
\begin{align}
 f(x,y_1,y_2,e^{i\alpha_i},e^{i\tilde\alpha_i}) =&
 -\sum_{i\ne j}x^{|n_i-n_j|}e^{-i(\alpha_i-\alpha_j)}
 +\sum_{i,j}f^+(x,y_1,y_2)x^{|n_i-\tilde n_j|}e^{-i(\alpha_i-\tilde\alpha_j)}
 \nonumber \\ &
 -\sum_{i\ne j}x^{|\tilde n_i-\tilde n_j|}e^{-i(\tilde\alpha_i-\tilde\alpha_j)}
 +\sum_{i,j}f^-(x,y_1,y_2)x^{|\tilde n_i-n_j|}e^{-i(\tilde\alpha_i-\alpha_j)},
 \nonumber \\
 f^+(x,y_1,y_2) =& \frac1{1-x^2}\left(
 x^\frac12y_1^{\frac12}y_2^{-\frac12}
+x^\frac12y_1^{-\frac12}y_2^{\frac12}
-x^\frac32y_1^{\frac12}y_2^{\frac12}
-x^\frac32y_1^{-\frac12}y_2^{-\frac12}\right),
 \nonumber \\
 f^-(x,y_1,y_2) =&  f^+(x,y_1,y_2^{-1})\,.
\end{align}
The Casimir energy is given by
\begin{equation}
 \epsilon_0 = \sum_{i,j}|n_i-\tilde n_j|-\sum_{i<j}|n_i-n_j|-\sum_{i<j}|\tilde n_i-\tilde n_j|\,.
\end{equation}
The full superconformal index $I(x,y_1,y_2,y_3)$ is given by the sum of (\ref{In}) over different flux sectors with an additional weight $y_3^{\frac k2\sum_in_i}$. The new fugacity parameter $y_3$ counts the KK momentum along the M-theory circle, i.e. Hopf fiber circle of $S^7/\mathbb Z_k$.

\paragraph{The large $N$ limit.}

A nice way \cite{Sundborg:1999ue,Aharony:2003sx} to treat the integral over the $N+N$ variables $\alpha_i,\tilde\alpha_i$ in the limit is to express it in terms of the eigenvalue density functions $\rho(\alpha),\tilde\rho(\tilde\alpha)$ and their Fourier modes,
\begin{alignat}{2}
 \rho(\alpha)&\equiv\sum_{i=1}^N\delta(\alpha-\alpha_i)
=\frac1{2\pi}\sum_{n\in\mathbb Z}\rho_ne^{in\alpha},\qquad&
 \rho_n&\equiv \sum_{i=1}^Ne^{-in\alpha_i},
 \nonumber \\
\tilde\rho(\tilde\alpha)&\equiv\sum_{i=1}^N\delta(\tilde\alpha-\tilde\alpha_i)
=\frac1{2\pi}\sum_{n\in\mathbb Z}\tilde\rho_ne^{in\tilde\alpha},\qquad&
 \tilde\rho_n&\equiv \sum_{i=1}^Ne^{-in\tilde\alpha_i}.
\end{alignat}
Note that $\rho_0=\tilde\rho_0=N$. As a simple exercise, let us rewrite the contribution of zero-flux sector $I^{(0)}$ using these variables. We find that the result is a simple Gaussian integral,
\begin{align}
 I^{(0)} =& \int\prod_{n\ne0}
 \mathrm d\rho_n\mathrm d\tilde\rho_n
 \cdot\exp\bigg(\sum_{n\ge1}\frac1nf(\cdot^n)\bigg),
 \nonumber \\ &
 f(\cdot^n) = -(\rho_n~~\tilde\rho_n)
\left(\begin{array}{cc} 1&-f^+(\cdot^n)\\-f^-(\cdot^n)&1\end{array}\right)
\left(\begin{array}{c}\rho_{-n}\\\tilde\rho_{-n}\end{array}\right),
\end{align}
which gives
\begin{equation}
 I^{(0)} = \prod_{n=1}^\infty\frac{(1-x^{2n})^2}
{(1-x^ny_1^n)(1-x^ny_1^{-n})(1-x^ny_2^n)(1-x^ny_2^{-n})}\,.
\end{equation}

The evaluation of contributions of the sectors with nonzero flux is apparently much harder. The idea employed in \cite{Kim:2009wb} is to divide the integration variables $\alpha_i,\tilde\alpha_i$ into 3 groups. The first contains those $\alpha_i$ or $\tilde\alpha_i$ for which the corresponding flux ($n_i$ or $\tilde n_i$) is positive, and the second contains those corresponding to negative flux. As long as one looks at the sectors carrying ${\cal O}(N^0)$ momentum along the M-theory circle, these two groups have ${\cal O}(N^0)$ variables. All the rest, corresponding to zero flux, are in the third group. In the large $N$ limit one can apply the change of variables described in the previous paragraph to the third group, after which the integration measure becomes schematically as follows.
\begin{equation}
 \frac1{\text{Sym}}\int\textrm d(\alpha,\tilde\alpha)~\Longrightarrow~~
 \frac1{\text{Sym}'}\int\textrm d(\alpha,\tilde\alpha)_+\,\textrm d(\alpha,\tilde\alpha)_-\prod_{n\ne0}\textrm d\rho_n\textrm d\tilde\rho_n\,.
\end{equation}
A nice observation of \cite{Kim:2009wb} is that the integral over $\rho_n,\tilde\rho_n$ is still Gaussian, and moreover the result takes the following factorized form.
\begin{equation}
 I\Big|_{(n_i,\tilde n_i)} = I^{(0)}\int\textrm d(\alpha,\tilde\alpha)_+\textrm d(\alpha,\tilde\alpha)_-\Big(\text{function of }(\alpha,\tilde\alpha)_+\Big)\cdot\Big(\text{function of }(\alpha,\tilde\alpha)_-\Big).
\end{equation}
This implies that the full index $I$ takes factorized form,
\begin{equation}
 I(x,y_1,y_2,y_3) = I^{(0)}(x,y_1,y_2)\cdot I_+(x,y_1,y_2,y_3)\cdot I_-(x,y_1,y_2,y_3),
\end{equation}
where $I_+,I_-$ are positive and negative power series in $y_3$, respectively. Though we are left with a finite-dimensional integral, the computation becomes increasingly complicated as the flux increases.

\section{Free energy on $S^3$}

Free energy measures the number of low-energy degrees of freedom. A supergravity analysis predicted \cite{Klebanov:1996un} that the free energy for the system of $N$ M2-branes should scale as $N^{3/2}$ at large $N$. This behavior was reproduced from the exact partition function of the ABJM model on $S^3$.

Generally, for a system of $N$ M2-branes with near-horizon geometry $\text{AdS}_4\times Y$, the gravitational free energy is given by the classical action evaluated on the corresponding Euclidean background. Though it is naively infinite, after subtracting the power-law divergences by suitable counterterms \cite{Henningson:1998gx,Emparan:1999pm,Balasubramanian:1999re} one can extract a finite positive value
\begin{equation}
 F = \frac{\pi L^2}{2G_{(4)}}.
\label{FSG0}
\end{equation}
Here $L$ is the radius of $\text{AdS}_4$ and $G_{(4)}$ is the effective 4D Newton constant. As a function of $N$ and the volume of $Y$ (normalized so that its metric satisfies $R_{mn}=6g_{mn}$), $F$ becomes
\begin{equation}
 F =  N^{\frac32}\sqrt{\frac{2\pi^6}{27\text{vol}(Y)}}\,.
\label{FSG}
\end{equation}
For the ABJM model one has $Y=S^7/\mathbb Z_k$ and $F=\frac{\sqrt2\pi}3 k^{\frac12}N^{\frac32}$.

The path integral of the Euclidean ABJM model on $S^3$ was studied in \cite{Kapustin:2009kz}. It was shown that the saddle points are parametrized by $2N$ variables $\sigma_i,\tilde\sigma_i~(i=1,\cdots,N)$, and the scalar fields in the vectormultiplets take constant values
\begin{equation}
 \sigma=\text{diag}(\sigma_1,\cdots,\sigma_N),\qquad
 \tilde\sigma=\text{diag}(\tilde\sigma_1,\cdots,\tilde\sigma_N)
\end{equation}
at the saddle points. The partition function is given by the following integral,
\begin{align}
 Z_{S^3} =& \int\mathrm d^N\sigma\mathrm d^N\tilde\sigma e^{-F(\sigma_i,\tilde\sigma_i)},
\nonumber \\
 F(\sigma_i,\tilde\sigma_i) =& -\frac{1}{2g_s}\sum_{i=1}^N\left(\sigma_i^2-\tilde\sigma_i^2\right)
-2\sum_{i<j}^N\ln\Big(2\sinh\frac{\sigma_i-\sigma_j}2\Big)
-2\sum_{i<j}^N\ln\Big( 2\sinh\frac{\tilde\sigma_i-\tilde\sigma_j}2\Big)
\nonumber \\ &+2\sum_{i,j=1}^N\ln\Big(2\cosh\frac{\sigma_i-\tilde\sigma_j}2\Big)
+2\ln N!+2N\ln(2\pi),
\label{ZS3}
\end{align}
where we introduced $g_s\equiv 2\pi i/k$.

\paragraph{Large $N$ limit: traditional approach.}

A standard way to evaluate this integral is to use the idea of large $N$ expansion \cite{Drukker:2010nc}. Let us generalize the gauge group to $U(N_1)\times U(N_2)$ for a while and consider the limit $N_1,N_2,k\to \infty$ with the 't Hooft couplings $t_1\equiv g_sN_1$ and $t_2\equiv g_sN_2$ kept fixed. The free energy then has an expansion of the form
\begin{equation}
 F = -\ln Z_{S^3} = \sum_{g=0}^\infty g_s^{2g-2}F_g(t_1,t_2).
\label{Fgdef}
\end{equation}
The planar contribution $g_s^{-2}F_0(t_1,t_2)$ is given by the value of $F(\sigma_i,\tilde\sigma_i)$ at its extremum where $\sigma_i,\tilde\sigma_i$ satisfy
\begin{align}
 0 &\;=\; \frac{\partial F}{\partial\sigma_i} \;=\; \phantom{+}\frac{\sigma_i}{g_s}-\sum_{j\ne i}\coth\frac{\sigma_i-\sigma_j}2+\sum_j\tanh\frac{\sigma_i-\tilde\sigma_j}2,
\nonumber \\
 0 &\;=\; \frac{\partial F}{\partial\tilde\sigma_i} \;=\; -\frac{\tilde\sigma_i}{g_s}-\sum_{j\ne i}\coth\frac{\tilde\sigma_i-\tilde\sigma_j}2+\sum_j\tanh\frac{\tilde\sigma_i-\sigma_j}2.
\label{eomev}
\end{align}
These equations are often interpreted as the condition for the equilibrium of forces acting on each eigenvalue. The forces between two $\sigma$'s or two $\tilde\sigma_i$'s are repulsive, whereas $\sigma_i$ and $\tilde\sigma_j$ attract each other.

\begin{figure}[t]
\begin{center}
\begin{tabular}{cc}
\begin{overpic}[scale=1]{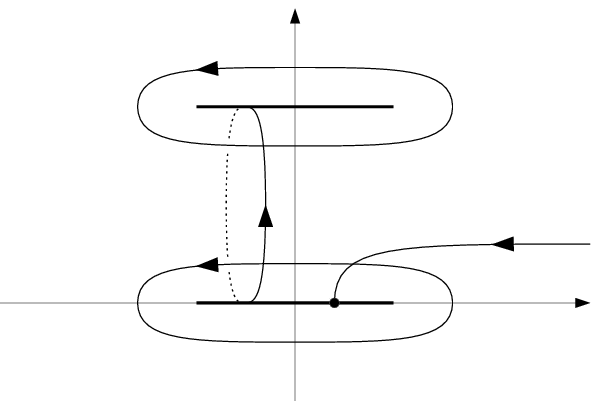}
\put(101,15){\scriptsize $\mathrm{Re}(z)$}
\put(47,68 ){\scriptsize $\mathrm{Im}(z)$}
\put(64,18){\scriptsize $C$}
\put(58,51.3){\scriptsize $\tilde C+i\pi$}
\put(55,12.5){\scriptsize $z_\ast$}
\put(33,25){\scriptsize $\alpha$}
\put(33,58){\scriptsize $\tilde\alpha$}
\put(46,32 ){\scriptsize $\beta$}
\put(83,29){\scriptsize $\gamma$}
\end{overpic}
\hskip10mm &\hskip10mm 
\begin{overpic}[scale=1]{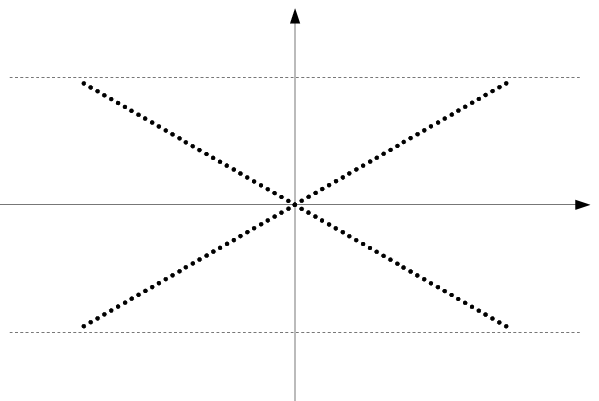}
\put(101,32){\scriptsize $\mathrm{Re}(\sigma)$}
\put(47,68 ){\scriptsize $\mathrm{Im}(\sigma)$}
\put(41,56 ){\scriptsize $\pi/2$}
\put(37,13 ){\scriptsize $-\pi/2$}
\end{overpic}
\end{tabular}
\end{center}
\caption{(Left) the branch cuts of $\omega(z)$ and the contours $\alpha,\tilde\alpha,\beta,\gamma$. (Right) a sketch of the numerical solution of (\ref{eomev}) found in \cite{Herzog:2010hf}.}\label{fig:contours}
\end{figure}

In the large $N$ limit, the eigenvalues $\{\sigma_i\}$ and $\{\tilde\sigma_i\}$ will form continuous distributions along some intervals $C$ and $\tilde C$. Let $\rho(x)$ and $\tilde\rho(x)$ be their densities. Moreover, let us define the resolvent by
\begin{align}
 \omega(z) &\;=\;
 g_s\sum_{j=1}^{N_1}\coth\frac{z-\sigma_j}2
-g_s\sum_{j=1}^{N_2}\tanh\frac{z-\tilde\sigma_j}2 \nonumber \\
 &\;=\; 
 t_1\int_C\mathrm dx\rho(x)\coth\frac{z-x}{2}
-t_2\int_{\tilde C}\mathrm dx\tilde\rho(x)\tanh\frac{z-x}2.
\end{align}
It turns out that the equations (\ref{eomev}) translate into the following discontinuity relations for $\omega(z)$,
\begin{align}
 \omega(z-i\epsilon)+\omega(z+i\epsilon)&=2z,\quad(\text{for}~~z\in C)\nonumber\\
 \omega(z+i\pi-i\epsilon)+\omega(z+i\pi+i\epsilon)&=2z.\quad(\text{for}~~z\in\tilde C)
\end{align}
which implies that $f(z)\equiv e^{\omega(z)}+e^{2z-\omega(z)}$ is an entire function. By combining it with the boundary condition at infinity one can determines $\omega(z)$ up to an arbitrary constant $\kappa$,
\begin{equation}
\omega=t_2-t_1+2\ln\frac12
\left(\sqrt{1+(i\kappa-2e^{t_1-t_2})e^z+e^{2z}}
     -\sqrt{1+(i\kappa+2e^{t_1-t_2})e^z+e^{2z}}\right).
\end{equation}
The square-roots produce two branch cuts $C$ and $\tilde C+i\pi$, and the left of Fig. \ref{fig:contours} shows their form when $t_1-t_2>0$ for a suitable choice of $\kappa$. A useful fact is that the $\kappa$-derivative of the integral $\int\omega(z)\mathrm dz$ is an elliptic integral,
\begin{equation}
 \frac\partial{\partial\kappa}\int\omega(z)\mathrm dz = \int\frac{-i \mathrm du}
{\sqrt{(1+(i\kappa-2e^{t_1-t_2})u+u^2)(1+(i\kappa+2e^{t_1-t_2})u+u^2)}},
\end{equation}
where we denoted $e^z\equiv u$.

To find a relation between the planar free energy and 't Hooft couplings, we express them using contour integrals of $\omega(z)\mathrm dz$. First of all, one finds
\begin{equation}
 \oint_\alpha \omega(z)\mathrm dz = 4\pi it_1,\qquad
 \oint_{\tilde\alpha} \omega(z)\mathrm dz = -4\pi it_2,
\end{equation}
where $\alpha,\tilde\alpha$ are the contours encircling $C$ and $\tilde C+i\pi$ as shown in Fig.\ref{fig:contours}. Also, notice that one can transport one $\sigma$ eigenvalue from infinity to a point $z_\ast\in C$ by integrating $\frac1{g_s}(z-\omega(z))$ along a contour $\gamma$ shown in the figure. If the integral were not divergent, it would correspond to the change of the free energy under the shift of $N_1$ by one (or the shift of $t_1$ by $g_s$). It turns out that a simultaneous shift of $t_1,t_2$ corresponds to a finite contour integral,
\begin{equation}
 \frac12\oint_\beta\omega(z)\mathrm dz = \frac{\partial F_0}{\partial t_1}+\frac{\partial F_0}{\partial t_2}+i\pi(t_1-t_2),
\end{equation}
where the contour $\beta$ is as shown in the figure.

We now restrict to the ABJM model and set $t_1=t_2\equiv t$. For a large positive $\kappa$, the four branch points of the elliptic integral are approximately at
\[
u\;\simeq\; -i\kappa+2,\quad\frac i\kappa+\frac 2{\kappa^2},\quad
\frac i\kappa-\frac 2{\kappa^2},\quad -i\kappa-2,
\]
with $C$ runnning between the first two and $\tilde C+i\pi$ between the latter two. It is not difficult to extract, by evaluating the elliptic integrals and integrating with respect to $\kappa$, the leading large $\kappa$ behavior
\begin{equation}
 t\sim \frac i\pi(\ln\kappa)^2,\qquad
 \frac{\partial F_0}{\partial t} \sim i\pi\ln\kappa.
\end{equation}
This implies $F_0\sim -\frac23(-i\pi t)^{3/2}$ and therefore $F\sim \frac{\sqrt2\pi}3k^{\frac12}N^{\frac32}$, which is in precise agreement with the prediction of supergravity.

The original work \cite{Drukker:2010nc} and \cite{Drukker:2011zy} also studied non-planar corrections (higher orders of perturbative series in $g_s$) and instanton corrections by making use of the connection of the integral (\ref{ZS3}) with the one for Chern-Simons theory on the lens space, which is in turn dual at large $N$ to  topological string theory on local $\mathbb P^1\times\mathbb P^1$. The full perturbative series (\ref{Fgdef}) was computed in \cite{Fuji:2011km} using holomorphic anomaly equation, and the result turned out to be given simply by  Airy function.

\paragraph{Large $N$ limit: another approach.}

A different method for evaluating the integral (\ref{ZS3}) was invented in \cite{Herzog:2010hf}, and it turned out very efficient for studying the large $N$ limit with $k$ fixed. It is partly based on the numerical result for the extrimization of $F(\sigma_i,\tilde\sigma_i)$ which look like the right of Fig.\ref{fig:contours}. It implied that the eigenvalue distribution is described by two functions $\rho(x),y(x)$ in such a way that the following replacement
\begin{align}
 \sum_{i=1}^N\varphi(\sigma_i) &\;\rightarrow\;
 N\int\mathrm dx\rho(x)\varphi\big(N^\alpha x+iy(x)\big),\nonumber\\
 \sum_{i=1}^N\varphi(\tilde\sigma_i) &\;\rightarrow\;
 N\int\mathrm dx\rho(x)\varphi\big(N^\alpha x-iy(x)\big)
\label{ansatz}
\end{align}
works for arbitrary function $\varphi(x)$ in the large $N$ limit. By rewriting $F(\sigma_i,\tilde\sigma_i)$ using this rule one finds it becomes a local functional of $\rho(x)$ and $y(x)$,
\begin{equation}
 F[\rho,y] \;=\; \frac k{\pi}N^{1+\alpha}\int\mathrm dx\,x\rho(x)y(x)
 + N^{2-\alpha}\int\mathrm dx\rho(x)^2\,f(2y(x)),
\end{equation}
where $f(x)$ is a function of period $2\pi$ and $f(x)=\pi^2-x^2$ for $|x|\le\pi$. The balance of the two terms in the right hand side requires $\alpha=\frac12$, which immediately implies the $N^{3/2}$ scaling of the free energy. The initial assumption that the distributions of $\text{Re}(\sigma_i)$ and $\text{Re}(\tilde\sigma_i)$ are described by the same function $\rho(x)$ is also justified, because otherwise there would be terms of higher order in $N$ remaining on the right hand side. It is now easy to extremize $F$ with respect to $\rho(x), y(x)$ under the condition $\int\mathrm dx\rho(x)=1$. The result reads
\begin{equation}
 \rho(x)=\frac1{2x_\ast} ~~(|x|\le x_\ast),\qquad
 x_\ast=\pi\sqrt{\frac 2k},\qquad
 y(x)=\frac{\pi x}{2x_\ast}.
\end{equation}
The value of $F[\rho,y]$ at this extremum is $F=\frac{\sqrt2\pi}3k^{\frac12}N^{\frac32}$, thus the supergravity result was correctly reproduced again.

Though this method is efficient, it is not very obvious how to go beyond the strict large $N$ limit. Another powerful method, called ``fermi gas'' approach, to study the model systematically at large $N$ with $k$ kept fixed was introduced in \cite{Marino:2011eh}. It is based on a reformulation of the integral (\ref{ZS3}) as the partition function of a 1D gas of $N$ non-interacting fermions with a non-trivial Hamiltonian. Combination of this approach with other methods from TBA and topological strings led to a very detailed understanding on the structure of non-perturbative corrections \cite{Putrov:2012zi,Hatsuda:2012hm,Hatsuda:2012dt,Hatsuda:2013gj,Calvo:2012du,Hatsuda:2013oxa}.

\paragraph{Generalization.}

The check of AdS/CFT via comparison of free energy on $S^3$ can be generalized to the cases with less SUSY. Explicit formula is known for the free energy of general ${\cal N}\ge2$ Chern-Simons matter theories \cite{Jafferis:2010un,Hama:2010av}. For $N$ M2-branes at the tip of some Calabi-Yau 4-fold cone $X$, the worldvolume dynamics is described by $U(N)^p$ CS-matter theories with matters satisfying certain condition. By adopting an ansatz similar to (\ref{ansatz}), one can show that the free energy for such theories scale as $N^{3/2}$ and obtain a local functional of $\rho(x)$ and $y_1(x),\cdots,y_p(x)$ \cite{Jafferis:2011zi}.

A new issue arises from the fact that, for general ${\cal N}=2$ theories of vector and chiral multiplets, the Lagrangian on $S^3$ at the starting point has arbitrariness in the assignment of R-charges to chiral multiplets. By extremizing the functional of $\rho(x),y_a(x)$ one therefore ends up with a function of the matter R-charges. As was proposed in \cite{Jafferis:2010un,Jafferis:2011zi} and proved in \cite{Closset:2012vg}, the correct assignment corresponding to the R-charge of ${\cal N}=2$ superconformal symmetry is the one which maximizes the free energy. As an illustrative exercise, let us break the SUSY of the ABJM model to ${\cal N}=2$ by assigning arbitrary R-charges $\Delta_1,\Delta_2,\Delta_3,\Delta_4$ to the chiral fields $A_1,A_2,B_3,B_4$, with a constraint
\begin{equation}
\sum_{a=1}^4\Delta_a=2\,.
\label{marg} 
\end{equation}
We also turn on the R-charge $\Delta_m$ for the momopole operator $T$ carrying a unit flux. By deriving the free energy functional and extremizing it with respect to $\rho(x),y(x)$, one obtains
\begin{equation}
 F \;=\; \frac{N^{\frac32}4\sqrt2\pi}{3k^{\frac32}}\sqrt
{(k\Delta_1-\Delta_m)(k\Delta_2-\Delta_m)(k\Delta_3+\Delta_m)(k\Delta_4+\Delta_m)}.
\label{FD}
\end{equation}
It has a flat direction under which $(\Delta_1,\cdots,\Delta_4,\Delta_m)$ shifts by $(\delta,\delta,-\delta,-\delta,k\delta)$, which is a reflection of the fact that the operators $A_1,A_2,B_3,B_4,T$ carry $U(1)$ gauge charges corresponding to $\text{Tr}(A_\mu-\tilde A_\mu)$. By extremizing with respect to the other non-flat directions with the constraint (\ref{marg}) one recovers $F=\frac{\sqrt2\pi}3k^{\frac12}N^\frac32$ again.

On the gravity side, we need to compute the volume of the 7D Sasaki-Einstein space $Y$ which is the base of the cone $X$. If $X$ is toric, there is a useful technique to compute the volume of $Y$ as a solution to a minimization problem \cite{Martelli:2005tp}. By definition $X$ has $U(1)^4$ symmetry, and one can regard $X$ as a $T^4$ fibration over a convex polyhedral cone ${\cal C}$ inside $\mathbb R^4$. Its K\"ahler form is given by
\begin{equation}
 \omega = \sum_{i=0}^3\mathrm dx_i\wedge\mathrm d\varphi_i,
\end{equation}
with $x_i$ the coordinages on the base and $\varphi_i\sim\varphi_i+2\pi$ on the fiber. We denote by $\vec v_a$ the inward-pointing normal vector to the $a$-th facet of ${\cal C}$. CY condition implies one may assume $v_{a0}=1$ for all $a$. Note that its components $v_{ai}$ are all integers since $\vec v_a$ also specifies the 1-cycle of $T^4$ which shrinks above the $a$-th facet.

There is a distinguished isometry $\sum_ib_i\partial_{\varphi_i}$, called Reeb vector, which is paired up with the radial vector field under a chosen complex structure of $X$. As was shown in \cite{Martelli:2005tp}, $\vec b$ contains some information on the (K\"ahler but not necessarily Ricci-flat) metric of $X$ which is actually enough to determine the volume of $Y$ and all of its 5-cycles. Consider a hyperplane
\begin{equation}
 \sum_{i=1}^3b_ix_i =\frac12,
\end{equation}
which intersects ${\cal C}$ to make a finite polytope $\Delta_b$. Then
\begin{equation}
 \text{vol}(Y) = 128\pi^4\text{vol}(\Delta_b).
\end{equation}
$Y$ is Sassaki-Einstein when $\vec b$ is chosen to minimize $\text{vol}(\Delta_b)$ under a condition $b_0=4$.

As an example, for $X=\mathbb C^4/\mathbb Z_k$ one can take
\begin{equation}
\def\arraystretch{.8}
(\vec v_1,\vec v_2,\vec v_3,\vec v_4)=
\left(\begin{array}{cccc} 1&1&1&1\\0&1&0&1\\0&0&1&1\\0&0&0&k\end{array}\right),~~
\vec   b=2\cdot\left(\begin{array}{c} 2\\\Delta_1+\Delta_3\\\Delta_1+\Delta_4\\k\Delta_1-\Delta_m\end{array}\right).
\def\arraystretch{1}
\end{equation}
The parametrization of $\vec b$ by $\Delta$'s can be obtained by matching the R-charges of M5-branes wrapping various 5-cycles with those of gauge invariant operators in ABJM. One then finds
\begin{equation}
 \text{vol}(Y) = \frac{\pi^4k^3}{48(k\Delta_1-\Delta_m)(k\Delta_2-\Delta_m)(k\Delta_3+\Delta_m)(k\Delta_4+\Delta_m)},
\end{equation}
where we used (\ref{marg}). Note that the above result reproduces (\ref{FD}) via (\ref{FSG}) before extremization. Generalization of this correspondence was studied in \cite{Amariti:2012tj,Lee:2014rca}, though it is not simple because the numbers of parameters on the gauge and gravity sides do not agree in general.

\section{Entropy of charged black holes}

According to AdS/CFT, any classical solution with AdS asymptotics should be described as an ensemble of states in the corresponding CFT. Construction of black holes in AdS spacetime was known to be considerably harder than those in flat spacetime, but an analytic solution for asymptotically $\text{AdS}_4$ static BPS black holes with magnetic (and electric) charges was found in \cite{Cacciatori:2009iz}. A natural question is whether the dual CFTs correctly accounts for their entropy as the degeneracy of states.

\paragraph{Black hole solutions and their entropy.}

The black hole solutions were found in 4D ${\cal N}=2$ supergravity with $n$ abelian vectormultiplets and gauging \cite{Andrianopoli:1996cm}. The bosonic fields in this theory are the metric $g_{\mu\nu}(x)$, $(n+1)$ gauge fields $A_\mu^\Lambda(x)~(\Lambda=0,\cdots,n)$ and $n$ complex scalars $z^i(x)~(i=1,\cdots,n)$ which parametrize a special K\"ahler manifold ${\cal M}$. There is a rank $2n+2$ holomorphic vector bundle over ${\cal M}$, and the K\"ahler potential of ${\cal M}$ is expressed in terms of its section $\Omega \equiv (X^\Lambda(z),{\cal F}_\Lambda(z))$ and its conjugate $\bar\Omega\equiv(\bar X^\Lambda(\bar z),\bar{\cal F}_\Lambda(\bar z))$ as
\begin{equation}
 {\cal K}(z,\bar z) = -\ln\big(i\langle\Omega,\bar\Omega\rangle\big),
\end{equation}
where $\langle\Omega,\bar\Omega\rangle\equiv {\cal F}_\Lambda\bar X^\Lambda-X^\Lambda\bar{\cal F}_\Lambda$ is the duality-invariant bilinear product. The covariant derivatives of $\Omega,\bar\Omega$ with respect to $z^i,\bar z^{\bar\imath}$ are
\begin{alignat}{2}
\nabla_i\Omega\equiv\partial_i\Omega+\partial_i{\cal K}\cdot\Omega
 &=(\nabla_iX^\Lambda, \nabla_i{\cal F}_\Lambda),\quad&
\nabla_{\bar\imath}\Omega \equiv \bar\partial_{\bar\imath}\Omega&=0,
 \nonumber \\
\nabla_{\bar\imath}\bar\Omega\equiv\partial_{\bar\imath}\bar\Omega
+\partial_{\bar\imath}{\cal K}\cdot\bar\Omega
 &=(\nabla_{\bar\imath}\bar X^\Lambda, \nabla_{\bar\imath}\bar{\cal F}_\Lambda),\quad&
\nabla_i\bar\Omega \equiv \partial_i\bar\Omega &=0,
\end{alignat}
and the K\"ahler metric on ${\cal M}$ is
\begin{equation}
g_{i\bar\jmath}=-\frac{\langle\nabla_i\Omega,\nabla_{\bar\jmath}\bar\Omega\rangle}{\langle\Omega,\bar\Omega\rangle}.
\end{equation}
The condition $\langle\Omega,\nabla_i\Omega\rangle=0$ implies the existence of the prepotential ${\cal F}(X)$, which is a homogeneous function of degree 2 in $X^\Lambda$ satisfying ${\cal F}_\Lambda(z)=\frac{\partial{\cal F}}{\partial X^\Lambda}(X(z))$. It also implies there is a symmetric matrix ${\cal N}_{\Lambda\Sigma}(z,\bar z)$ such that
\begin{equation}
 F_\Lambda= {\cal N}_{\Lambda\Sigma}X^\Sigma,
\quad
 \nabla_{\bar\imath}\bar F_\Lambda= {\cal N}_{\Lambda\Sigma}\nabla_{\bar\imath}\bar X^\Sigma.
\end{equation}

The first few terms in the supergravity action \cite{Ferrara:1996dd} reads
\begin{equation}
 S = \frac1{16\pi G_\text{N}}\int\left(-\mathrm d\text{vol}\cdot R
 +{\cal N}_{\Lambda\Sigma}F^{+\Lambda}\wedge F^{+\Sigma}
 +\bar{\cal N}_{\Lambda\Sigma}F^{-\Lambda}\wedge F^{-\Sigma}
+\cdots\right),
\end{equation}
where $F^{\pm\Lambda}$ is the imaginary (anti-)self-dual part of the field strengths $F^\Lambda=\mathrm dA^\Lambda$  satisfying $\ast F^{\pm\Lambda}=\pm iF^{\pm\Lambda}$. We define the magnetic and electric charges $q\equiv (q^\Lambda,q_\Lambda)$ of spherically symmetric solutions by
\begin{equation}
 q^\Lambda\equiv\int_{S^2}\frac{F^\Lambda}{4\pi},\quad
 q_\Lambda\equiv\int_{S^2}\frac{G_\Lambda}{4\pi}.\qquad
\Big(G_\Lambda\equiv {\cal N}_{\Lambda\Sigma}F^{+\Sigma}+\bar{\cal N}_{\Lambda\Sigma}F^{-\Sigma}\Big)
\end{equation}
The duality group $Sp(2n+2,\mathbb R)$ rotates the vectors $\Omega$ and $q$ in the same way.

Static extremal black holes with flat asymptotics was studied in \cite{Ferrara:1996dd}. It was found that, for spherically symmetric solutions of the form
\begin{equation}
 \mathrm ds^2 = e^{2U(r)}\mathrm dt^2-e^{-2U(r)}\left\{
 \mathrm dr^2+r^2(\mathrm d\theta^2+\sin^2\theta\mathrm d\varphi^2) \right\},\quad
 z^i=z^i(r),
\end{equation}
the BPS condition can be cast into a flow equation,
\begin{equation}
 r^2\frac{\mathrm dU}{\mathrm dr}=e^U|Z|,\quad
 r^2\frac{\mathrm dz^i}{\mathrm dr}=2e^Ug^{i\bar\jmath}\frac\partial{\partial\bar z^{\bar\jmath}}|Z|.
\label{flow1}
\end{equation}
Here $g^{i\bar\jmath}(z,\bar z)$ is the inverse metric on ${\cal M}$ and $Z(z,\bar z)\equiv e^{{\cal K}/2}\langle\Omega,q\rangle$ is the central charge of the black hole with charge $q$. These imply that the value of the scalars $z^i$ at the horizon $r=0$ should extremize $|Z(z,\bar z)|$, and also that the entropy of the black hole is given by
\begin{equation}
 S_\text{BH} = \frac1{4G_\text{N}}\text{(horizon area)} = \frac{\pi|Z|^2}{G_\text{N}}
\end{equation}
at its extremum, which is therefore a function of the charge $q$ only.

To discuss black holes with AdS asymptotics, one needs to move to gauged supergravity. In 4D ${\cal N}=2$ supergravity, gauging amounts to assigning $U(1)^{n+1}$ charges to fields according to their $SU(2)_\text{R}$ charges. We denote the couplings as $g\equiv(g^\Lambda,g_\Lambda)$, though the discussions of concrete theories are often restricted to those with $g^\Lambda=0$.

In order to explain the mechanism of gauging, we think of adding $n_\text{H}$ hypermultiplets whose scalars $y^m~(m=1,\cdots,4n_\text{H})$ parametrize a quaternionic space ${\cal M}_\text{H}$. There is a principal $SU(2)$ bundle over ${\cal M}_\text{H}$ with connection $V^a=V^a_m(y)\mathrm dy^m~(a=1,2,3)$ such that the triplet of K\"ahler forms of ${\cal M}_\text{H}$ is proportional to its curvature 2-form. All the fields with $SU(2)_\text{R}$ charges are then coupled to $V^a$. For example, the SUSY transformation rule for gravitino $\psi_\mu^A(x)$ reads
\begin{equation}
 \delta_\epsilon\psi^A_\mu = \partial_\mu\epsilon^A+\frac14\omega^{ab}_\mu\gamma^{ab}\epsilon^A-\frac i2\partial_\mu y^mV^a_m(y)(\sigma^a)^A_{~B}\epsilon^B+\cdots.
\end{equation}
If ${\cal M}_\text{H}$ has a $U(1)^{n+1}$ isometry and we want to gauge it, we covariantize the derivatives
\begin{equation}
 \partial_\mu y^m(x) \longrightarrow \partial_\mu y^m(x) + A^\Lambda_\mu(x)\cdot k^m_\Lambda(y(x)),
\end{equation}
where $k^m_\Lambda(y)$ is the Killing vector field on ${\cal M}_\text{H}$ for the $\Lambda$-th $U(1)$. At the same time, an analogue of the $U(1)^{n+1}$ hyperK\"ahler moment map $P^a_\Lambda(y)$ takes part in the modification
\begin{equation}
 \partial_\mu y^m(x)V^a_m(y(x)) \longrightarrow \partial_\mu y^m(x)V^a_m(y(x)) + A^\Lambda_\mu(x)\cdot P^a_\Lambda(y(x))\,.
\end{equation}
This procedure works even for the case with empty ${\cal M}_\text{H}$ and constant $P^a_\Lambda(y)=g_\Lambda\delta^{a3}$, and thus couples the $U(1)^{n+1}$ gauge fields to fields with $SU(2)_\text{R}$ charges. The gravitino SUSY transformation rule now becomes
\begin{equation}
 \delta_\epsilon(\psi^A_\mu\mathrm dx^\mu) = \mathrm d\epsilon^A+\frac14\omega^{ab}\gamma^{ab}\epsilon^A-\frac i2g_\Lambda A^\Lambda(\sigma^3)^A_{~B}\epsilon^B+\cdots.
\label{dpsi}
\end{equation}
Note that the coupling $g_\Lambda$ determines the quantization rule of the charges,
\begin{equation}
 q^\Lambda\in\frac1{2g_\Lambda}\,\mathbb Z,\qquad
 q_\Lambda \in 2g_\Lambda G_\text{N}\,\mathbb Z.
\end{equation}

A peculiar feature of the black hole solutions of \cite{Cacciatori:2009iz} with magnetic charge is that, with a spherical symmetric ansatz, the gauge field $A^\Lambda$ and spin connection $\omega^{ab}$ both take the form $\sim\cos\theta\mathrm d\varphi$ so that the second and the third terms in (\ref{dpsi}) cancel each other. This occurs when the twisting condition
\begin{equation}
 g_\Lambda q^\Lambda=-1
\label{twc}
\end{equation}
is satisfied. The cancellation among contributions from various connections is reminiscent of the construction of topologically twisted theories. The observation of this fact led to the identification of the dual description \cite{Benini:2015eyy,Benini:2016rke}. Black hole solutions with $g_\Lambda q^\Lambda=0$ are also known but they require some rotation to be free of naked singularity \cite{Cvetic:2005zi,Hristov:2019mqp}.

For BPS black hole solutions in gauged supergravity, there is an attractor flow equation similar to (\ref{flow1}) which allows one to obtain the black hole entropy without working out the metric explicitly \cite{DallAgata:2010ejj}. It implies that the value of $z^i$ at the horizon extremizes
\begin{equation}
 R^2 \equiv -i\frac{\langle q,\Omega\rangle}{\langle g,\Omega\rangle}
 = -i\frac{q_\Lambda X^\Lambda-q^\Lambda{\cal F}_\Lambda}{g_\Lambda X^\Lambda},
\end{equation}
and the value of $R$ at the extremum gives the horizon radius. Thanks to the homogeneity of the RHS, one can think of extremizing the numerator as a function of $X^\Lambda$ keeping the denominator fixed.

As an example, let us consider ${\cal N}=8$ maximal gauged supergravity truncated to ${\cal N}=2$ which is relevant to the ABJM model at $k=1$. Its prepotential and couplings are given by
\begin{equation}
 {\cal F}(X)=-2i\sqrt{X^0X^1X^2X^3},\qquad
 g_0=g_1=g_2=g_3\equiv g=\frac1{\sqrt2L},
\end{equation}
where $L$ is the $\text{AdS}_4$ radius. In terms of integer charges
\begin{equation}
 n^\Lambda\equiv -2gq^\Lambda,\qquad
 e_\Lambda\equiv \frac{q_\Lambda}{2gG_\text{N}},
\end{equation}
the twisting condition (\ref{twc}) becomes $\sum_\Lambda n^\Lambda=2$, and the entropy is given by
\begin{equation}
 S_\text{BH} = -i\bigg(e_\Lambda X^\Lambda + \frac{L^2}{2G_\text{N}}\cdot n^\Lambda\frac{\partial{\cal F}}{\partial X^\Lambda}\bigg),
\label{SBH}
\end{equation}
extremized as a function of $X^\Lambda$ under a constraint $\sum_\Lambda X^\Lambda=2\pi$. Note that the extremum value has to be real and positive in order for the solution to have a smooth horizon. This puts an independent condition on $(n^\Lambda,e_\Lambda)$.

\paragraph{Microscopic theory.}

The microscopic theory for these black holes is a 3D ${\cal N}=2$ supersymmetric theory on $S^2\times\mathbb R$ with a topological twist by a unit background $U(1)_\text{R}$ flux through $S^2$. The matter R-charge has to be integer due to Dirac quantization, but there are infinite choices for its assignments if the theory has flavor symmetry. The path integral of the theory with periodic time (i.e. on $S^1\times S^2$) is called the twisted index \cite{Benini:2015noa}. For the theories with conserved flavor charges $J^a$, one can turn on the constant $\sigma$ and $A_\tau$ components of the corresponding vector multiplet in the background. In Hamiltonian description, the twisted index computes the trace over the Hilbert space ${\cal H}$,
\begin{equation}
 I = \text{Tr}_{\cal H}\Big[(-1)^Fe^{-\beta(H-i\sum_aA_\tau^aJ^a)}\Big].
\end{equation}
It is a function of the complexified flat connections $\Delta_a\equiv \beta(A_\tau^a+i\sigma^a)$ only, because the supercharge $Q$ satisfies $Q^2=H-\sum_a\sigma^aJ^a$.

In the case of the ABJM model, the twist is labeled by the R-charges $n_1,\cdots,n_4\in\mathbb Z$ of the chiral multiplets $A_1,A_2,B_3,B_4$ obeying a constraint $\sum_an_a=2$. The model has $U(1)^3$ flavor symmetry generated by $J^a-J^4~(a=1,2,3)$, where $J^a$ phase-rotates the $a$-th chiral multiplet. It is therefore convenient to regard $I$ as a function of flat connections $\Delta_1,\cdots,\Delta_4$ obeying a constraint
\begin{equation}
 \Delta_1+\Delta_2+\Delta_3+\Delta_4=0~~~\text{mod}~~2\pi\mathbb Z.
\end{equation}
Thanks to SUSY localization, the path integral can be reduced to an integral over saddle points labeled by integers $m_i,\tilde m_i$ and periodic variables $u_i,\tilde u_i~(i=1,\cdots,N)$. They appear in the value of vectormultiplet field at the saddle point as follows,
\begin{alignat}{2}
 \text P\exp i\int_{S^1}(A+i\sigma\mathrm d\tau)&=\text{diag}(e^{iu_1},\cdots,e^{iu_N}),\qquad&
 \int_{S^2}\frac{F}{2\pi}&=\text{diag}(m_1,\cdots,m_N),
 \nonumber \\
 \text P\exp i\int_{S^1}(\tilde A+i\tilde\sigma\mathrm d\tau)&=\text{diag}(e^{i\tilde u_1},\cdots,e^{i\tilde u_N}),\qquad&
 \int_{S^2}\frac{\tilde F}{2\pi}&=\text{diag}(\tilde m_1,\cdots,\tilde m_N).
\end{alignat}
The system also has fermionic zeromodes $\xi_i,\tilde\xi_i$ which are paired with $u_i^\ast,\tilde u_i^\ast$ under the supersymmetry. Consequently, after localization one is left with an integral over $u$'s along some contour, not over the cylinder. With $x_i\equiv e^{iu_i}, \tilde x_i\equiv e^{i\tilde u_i}$ and $y_a\equiv e^{i\Delta_a}$ one can express the index as follows.
\begin{align}
 I &= \frac1{(N!)^2}\sum_{m_i,\tilde m_i}\int\prod_{i=1}^N
 \frac{\mathrm dx_i}{2\pi ix_i}
 \frac{\mathrm d\tilde x_i}{2\pi i\tilde x_i}
 x_i^{km_i}\tilde x_i^{-k\tilde m_i}\cdot
 \prod_{i\ne j}^N\Big(1-\frac{x_i}{x_j}\Big)
 \Big(1-\frac{\tilde x_i}{\tilde x_j}\Big)\nonumber \\ &\hskip26mm\cdot
 \prod_{i,j=1}^N\Bigg\{
 \prod_{a=1,2}\bigg(\frac{\sqrt{y_ax_i/\tilde x_j}}{1-y_ax_i/\tilde x_j}\bigg)^{m_i-\tilde m_j-n_a+1}
 \nonumber \\ & \hskip38mm\cdot
 \prod_{b=3,4}\bigg(\frac{\sqrt{y_b\tilde x_j/x_i}}{1-y_b\tilde x_j/x_i}\bigg)^{\tilde m_j-m_i-n_b+1}
 \Bigg\}\,.
\label{tind}
\end{align}

The contour integral is performed following the Jeffrey-Kirwan residue prescription \cite{Jeffrey:1993}, which was first used in the study of SUSY localized path integrals in 2D \cite{Benini:2013nda,Benini:2013xpa} and 1D \cite{Hori:2014tda}. It goes roughly as follows. For each pole (intersections of singular hyperplanes) $p$ of the integrand, there is a matter field responsible for each of the hyperplanes. Label $p$ by the charges $Q_p=\{\vec q_1,\vec q_2,\cdots\}$ of those matters under the Cartan of the gauge group. (In the present problem there are additional singularities at $x_i,\tilde x_i=0$ or $\infty$. They are labeled according to the Chern-Simons couplings \cite{Benini:2015noa}.) The JK-residue prescription begins by choosing a reference charge $\vec\eta$ arbitrarily. Then one decides whether to pick up the residue of a pole $p$ according to whether the cone spanned by the charge vectors in $Q_p$ includes $\vec\eta$ or not. The end result is independent of the initial choice of $\vec\eta$.

For the index of the ABJM model (\ref{tind}), there is a suitable choice of $\vec\eta$ such that one only has to evaluate the residue of the pole at $x_i=\tilde x_i=0$. Then the terms in $I$ with $m_i$ very large (or $\tilde m_i$ negatively very large) can be discarded because there would not be a pole at $x_i=0$ (or $\tilde x_i=0$). As a result, one only has to sum over $m_i\le M$ and $\tilde m_i\ge -M$ for some $M$, which can be performed easily before integrating over $x_i,\tilde x_i$. One is then left with an integral over $x_i,\tilde x_i$, and the integrand has poles at the solution of a Bethe ansatz like equations
\begin{equation}
 x_i^k = \prod_{j=1}^N\frac
{(1-\tilde x_j/x_iy_1)(1-\tilde x_j/x_iy_2)}
{(1-\tilde y_3x_j/x_i)(1-y_4\tilde x_j/x_i)},\quad
 \tilde x_j^k = \prod_{i=1}^N\frac
{(1-\tilde x_j/x_iy_1)(1-\tilde x_j/x_iy_2)}
{(1-\tilde y_3x_j/x_i)(1-y_4\tilde x_j/x_i)}.
\end{equation}
These are actually the equations for the extremum of the potential,
\begin{equation}
 \widetilde{W} \equiv \frac k2\sum_{i=1}^N(\tilde u_i^2-u_i^2)
 -\sum_{i,j=1}^N\sum_{a=1}^4\varepsilon_a
 \text{Li}_2\big(e^{i(\tilde u_j-u_i-\varepsilon_a\Delta_a)}\big)
 -2\pi\sum_{i=1}^N(\tilde c_i\tilde u_i-c_iu_i)\,.
\end{equation}
Here $\text{Li}_n(x)=\sum_{k\ge1}\frac{x^k}{k^n}$ is the polylogarithm function and $\varepsilon_a=(+1,+1,-1,-1)$. $c_i,\tilde c_i$ are integers which arise from the multi-valuedness of $\log$ function.

The extremization of $\widetilde{W}$ was studied in \cite{Benini:2015eyy}. It was found that by using an ansatz similar to (\ref{ansatz}),
\begin{align}
 \sum_{i=1}^N\varphi(u_i) &\;\to\;
 N\int\mathrm dx\rho(x)\varphi(iN^\alpha x+y(x)),\nonumber \\
 \sum_{i=1}^N\varphi(\tilde u_i) &\;\to\;
 N\int\mathrm dx\rho(x)\varphi(iN^\alpha x+\tilde y(x)),
\end{align}
one can rewrite $\widetilde W$ into a local functional of $\rho(x)$ and $\tilde y(x)-y(x)$, and moreover the local functional takes the same form as the one for the free energy (of the ${\cal N}=2$ deformed theory with general R-charge assignments) on $S^3$. It turned out that the variational problem has a consistent solution only for $\sum_a\Delta_a=2\pi$, and the value of $\widetilde W$ and $I$ at the solution are given (for $k=1$) by
\begin{equation}
 \ln I(n_a;\Delta_a) = i\sum_{a=1}^4n_a\frac{\partial\widetilde W}{\partial\Delta_a},\qquad
 \widetilde{W} = iN^{\frac32}\cdot\frac{2\sqrt2}3\sqrt{\Delta_1\Delta_2\Delta_3\Delta_4}\,.
\end{equation}
Now recall that $I(n_a;\Delta_a)$ is the trace over the states of the twisted theory labeled by $n_a$, with the weight $e^{i\sum_ae_a\Delta_a}$ for the states with flavor charge $J^a=e_a$. The number $d(n_a;e_a)$ of states with charge $e_a$ should therefore be related to $I$ by Legendre transformation,
\begin{equation}
 \ln d(n_a;e_a) = \sum_{a=1}^4\left(-ie_a\Delta_a+in_a\frac{\partial\widetilde W}{\partial\Delta_a}\right),
\end{equation}
where the RHS should be extremized as a function of $\Delta_a$ with a constraint $\sum_a\Delta_a=2\pi$. In view of (\ref{FSG0}), we see that the black hole entropy (\ref{SBH}) has been reproduced precisely by the microscopic theory.

~

Ever since I became a student of Prof.~Eguchi, I used to feel tense every time we had discussions of physics. It always led me to commit to physics more seriously and helped me grow. The same feeling comes back still now whenever I remember him.

I am truly grateful for being a student of Prof.~Tohru Eguchi.

~

\section*{Acknowledgments}

I would like to thank the organizers of the memorial conference ``particle physics and mathematical physics -- 40 years after Eguchi-Hanson solution'' in Kyoto. I would also like to thank the organizers of the workshop ``on challenges and advances in theoretical physics'' in Seoul where part of the content of this article was presented.

\bibliographystyle{utphys}
\bibliography{ref}

\providecommand{\href}[2]{#2}\begingroup\raggedright\begin{thebibliography}{10}

\bibitem{Maldacena:1997re}
J.~M. Maldacena, ``{The Large N limit of superconformal field theories and
  supergravity},'' \href{http://dx.doi.org/10.1023/A:1026654312961,
  10.4310/ATMP.1998.v2.n2.a1}{{\em Int. J. Theor. Phys.} {\bfseries 38} (1999)
  1113--1133}, \href{http://arxiv.org/abs/hep-th/9711200}{{\ttfamily
  arXiv:hep-th/9711200 [hep-th]}}.
[Adv. Theor. Math. Phys.2,231(1998)].

\bibitem{Witten:1998qj}
E.~Witten, ``{Anti-de Sitter space and holography},''
  \href{http://dx.doi.org/10.4310/ATMP.1998.v2.n2.a2}{{\em Adv. Theor. Math.
  Phys.} {\bfseries 2} (1998) 253--291},
\href{http://arxiv.org/abs/hep-th/9802150}{{\ttfamily arXiv:hep-th/9802150
  [hep-th]}}.

\bibitem{Hosomichi:1997if}
K.~Hosomichi, ``{Fermion emission from five-dimensional black holes},''
  \href{http://dx.doi.org/10.1016/S0550-3213(98)00192-8}{{\em Nucl. Phys.}
  {\bfseries B524} (1998) 312--332},
\href{http://arxiv.org/abs/hep-th/9711072}{{\ttfamily arXiv:hep-th/9711072
  [hep-th]}}.

\bibitem{Aharony:2008ug}
O.~Aharony, O.~Bergman, D.~L. Jafferis, and J.~Maldacena, ``{N=6 superconformal
  Chern-Simons-matter theories, M2-branes and their gravity duals},''
  \href{http://dx.doi.org/10.1088/1126-6708/2008/10/091}{{\em JHEP} {\bfseries
  10} (2008) 091},
\href{http://arxiv.org/abs/0806.1218}{{\ttfamily arXiv:0806.1218 [hep-th]}}.

\bibitem{Kinney:2005ej}
J.~Kinney, J.~M. Maldacena, S.~Minwalla, and S.~Raju, ``{An Index for 4
  dimensional super conformal theories},''
  \href{http://dx.doi.org/10.1007/s00220-007-0258-7}{{\em Commun. Math. Phys.}
  {\bfseries 275} (2007) 209--254},
\href{http://arxiv.org/abs/hep-th/0510251}{{\ttfamily arXiv:hep-th/0510251
  [hep-th]}}.

\bibitem{Bhattacharya:2008zy}
J.~Bhattacharya, S.~Bhattacharyya, S.~Minwalla, and S.~Raju, ``{Indices for
  Superconformal Field Theories in 3,5 and 6 Dimensions},''
  \href{http://dx.doi.org/10.1088/1126-6708/2008/02/064}{{\em JHEP} {\bfseries
  02} (2008) 064},
\href{http://arxiv.org/abs/0801.1435}{{\ttfamily arXiv:0801.1435 [hep-th]}}.

\bibitem{Bhattacharya:2008bja}
J.~Bhattacharya and S.~Minwalla, ``{Superconformal Indices for N = 6 Chern
  Simons Theories},''
  \href{http://dx.doi.org/10.1088/1126-6708/2009/01/014}{{\em JHEP} {\bfseries
  01} (2009) 014},
\href{http://arxiv.org/abs/0806.3251}{{\ttfamily arXiv:0806.3251 [hep-th]}}.

\bibitem{Kim:2009wb}
S.~Kim, ``{The Complete superconformal index for N=6 Chern-Simons theory},''
  \href{http://dx.doi.org/10.1016/j.nuclphysb.2012.07.015,
  10.1016/j.nuclphysb.2009.06.025}{{\em Nucl. Phys.} {\bfseries B821} (2009)
  241--284}, \href{http://arxiv.org/abs/0903.4172}{{\ttfamily arXiv:0903.4172
  [hep-th]}}.
[Erratum: Nucl. Phys.B864,884(2012)].

\bibitem{Sundborg:1999ue}
B.~Sundborg, ``{The Hagedorn transition, deconfinement and N=4 SYM theory},''
  \href{http://dx.doi.org/10.1016/S0550-3213(00)00044-4}{{\em Nucl. Phys.}
  {\bfseries B573} (2000) 349--363},
\href{http://arxiv.org/abs/hep-th/9908001}{{\ttfamily arXiv:hep-th/9908001
  [hep-th]}}.

\bibitem{Aharony:2003sx}
O.~Aharony, J.~Marsano, S.~Minwalla, K.~Papadodimas, and M.~Van~Raamsdonk,
  ``{The Hagedorn - deconfinement phase transition in weakly coupled large N
  gauge theories},'' \href{http://dx.doi.org/10.4310/ATMP.2004.v8.n4.a1}{{\em
  Adv. Theor. Math. Phys.} {\bfseries 8} (2004) 603--696},
  \href{http://arxiv.org/abs/hep-th/0310285}{{\ttfamily arXiv:hep-th/0310285
  [hep-th]}}.
[,161(2003)].

\bibitem{Klebanov:1996un}
I.~R. Klebanov and A.~A. Tseytlin, ``{Entropy of near extremal black
  p-branes},'' \href{http://dx.doi.org/10.1016/0550-3213(96)00295-7}{{\em Nucl.
  Phys.} {\bfseries B475} (1996) 164--178},
\href{http://arxiv.org/abs/hep-th/9604089}{{\ttfamily arXiv:hep-th/9604089
  [hep-th]}}.

\bibitem{Henningson:1998gx}
M.~Henningson and K.~Skenderis, ``{The Holographic Weyl anomaly},''
  \href{http://dx.doi.org/10.1088/1126-6708/1998/07/023}{{\em JHEP} {\bfseries
  07} (1998) 023},
\href{http://arxiv.org/abs/hep-th/9806087}{{\ttfamily arXiv:hep-th/9806087
  [hep-th]}}.

\bibitem{Emparan:1999pm}
R.~Emparan, C.~V. Johnson, and R.~C. Myers, ``{Surface terms as counterterms in
  the AdS / CFT correspondence},''
  \href{http://dx.doi.org/10.1103/PhysRevD.60.104001}{{\em Phys. Rev.}
  {\bfseries D60} (1999) 104001},
\href{http://arxiv.org/abs/hep-th/9903238}{{\ttfamily arXiv:hep-th/9903238
  [hep-th]}}.

\bibitem{Balasubramanian:1999re}
V.~Balasubramanian and P.~Kraus, ``{A Stress tensor for Anti-de Sitter
  gravity},'' \href{http://dx.doi.org/10.1007/s002200050764}{{\em Commun. Math.
  Phys.} {\bfseries 208} (1999) 413--428},
\href{http://arxiv.org/abs/hep-th/9902121}{{\ttfamily arXiv:hep-th/9902121
  [hep-th]}}.

\bibitem{Kapustin:2009kz}
A.~Kapustin, B.~Willett, and I.~Yaakov, ``{Exact Results for Wilson Loops in
  Superconformal Chern-Simons Theories with Matter},''
  \href{http://dx.doi.org/10.1007/JHEP03(2010)089}{{\em JHEP} {\bfseries 03}
  (2010) 089},
\href{http://arxiv.org/abs/0909.4559}{{\ttfamily arXiv:0909.4559 [hep-th]}}.

\bibitem{Drukker:2010nc}
N.~Drukker, M.~Marino, and P.~Putrov, ``{From weak to strong coupling in ABJM
  theory},'' \href{http://dx.doi.org/10.1007/s00220-011-1253-6}{{\em Commun.
  Math. Phys.} {\bfseries 306} (2011) 511--563},
\href{http://arxiv.org/abs/1007.3837}{{\ttfamily arXiv:1007.3837 [hep-th]}}.

\bibitem{Herzog:2010hf}
C.~P. Herzog, I.~R. Klebanov, S.~S. Pufu, and T.~Tesileanu, ``{Multi-Matrix
  Models and Tri-Sasaki Einstein Spaces},''
  \href{http://dx.doi.org/10.1103/PhysRevD.83.046001}{{\em Phys. Rev.}
  {\bfseries D83} (2011) 046001},
\href{http://arxiv.org/abs/1011.5487}{{\ttfamily arXiv:1011.5487 [hep-th]}}.

\bibitem{Drukker:2011zy}
N.~Drukker, M.~Marino, and P.~Putrov, ``{Nonperturbative aspects of ABJM
  theory},'' \href{http://dx.doi.org/10.1007/JHEP11(2011)141}{{\em JHEP}
  {\bfseries 11} (2011) 141},
\href{http://arxiv.org/abs/1103.4844}{{\ttfamily arXiv:1103.4844 [hep-th]}}.

\bibitem{Fuji:2011km}
H.~Fuji, S.~Hirano, and S.~Moriyama, ``{Summing Up All Genus Free Energy of
  ABJM Matrix Model},'' \href{http://dx.doi.org/10.1007/JHEP08(2011)001}{{\em
  JHEP} {\bfseries 08} (2011) 001},
\href{http://arxiv.org/abs/1106.4631}{{\ttfamily arXiv:1106.4631 [hep-th]}}.

\bibitem{Marino:2011eh}
M.~Marino and P.~Putrov, ``{ABJM theory as a Fermi gas},''
  \href{http://dx.doi.org/10.1088/1742-5468/2012/03/P03001}{{\em J. Stat.
  Mech.} {\bfseries 1203} (2012) P03001},
\href{http://arxiv.org/abs/1110.4066}{{\ttfamily arXiv:1110.4066 [hep-th]}}.

\bibitem{Putrov:2012zi}
P.~Putrov and M.~Yamazaki, ``{Exact ABJM Partition Function from TBA},''
  \href{http://dx.doi.org/10.1142/S0217732312502008}{{\em Mod. Phys. Lett.}
  {\bfseries A27} (2012) 1250200},
\href{http://arxiv.org/abs/1207.5066}{{\ttfamily arXiv:1207.5066 [hep-th]}}.

\bibitem{Hatsuda:2012hm}
Y.~Hatsuda, S.~Moriyama, and K.~Okuyama, ``{Exact Results on the ABJM Fermi
  Gas},'' \href{http://dx.doi.org/10.1007/JHEP10(2012)020}{{\em JHEP}
  {\bfseries 10} (2012) 020},
\href{http://arxiv.org/abs/1207.4283}{{\ttfamily arXiv:1207.4283 [hep-th]}}.

\bibitem{Hatsuda:2012dt}
Y.~Hatsuda, S.~Moriyama, and K.~Okuyama, ``{Instanton Effects in ABJM Theory
  from Fermi Gas Approach},''
  \href{http://dx.doi.org/10.1007/JHEP01(2013)158}{{\em JHEP} {\bfseries 01}
  (2013) 158},
\href{http://arxiv.org/abs/1211.1251}{{\ttfamily arXiv:1211.1251 [hep-th]}}.

\bibitem{Hatsuda:2013gj}
Y.~Hatsuda, S.~Moriyama, and K.~Okuyama, ``{Instanton Bound States in ABJM
  Theory},'' \href{http://dx.doi.org/10.1007/JHEP05(2013)054}{{\em JHEP}
  {\bfseries 05} (2013) 054},
\href{http://arxiv.org/abs/1301.5184}{{\ttfamily arXiv:1301.5184 [hep-th]}}.

\bibitem{Calvo:2012du}
F.~Calvo and M.~Marino, ``{Membrane instantons from a semiclassical TBA},''
  \href{http://dx.doi.org/10.1007/JHEP05(2013)006}{{\em JHEP} {\bfseries 05}
  (2013) 006},
\href{http://arxiv.org/abs/1212.5118}{{\ttfamily arXiv:1212.5118 [hep-th]}}.

\bibitem{Hatsuda:2013oxa}
Y.~Hatsuda, M.~Marino, S.~Moriyama, and K.~Okuyama, ``{Non-perturbative effects
  and the refined topological string},''
  \href{http://dx.doi.org/10.1007/JHEP09(2014)168}{{\em JHEP} {\bfseries 09}
  (2014) 168},
\href{http://arxiv.org/abs/1306.1734}{{\ttfamily arXiv:1306.1734 [hep-th]}}.

\bibitem{Jafferis:2010un}
D.~L. Jafferis, ``{The Exact Superconformal R-Symmetry Extremizes Z},''
  \href{http://dx.doi.org/10.1007/JHEP05(2012)159}{{\em JHEP} {\bfseries 05}
  (2012) 159},
\href{http://arxiv.org/abs/1012.3210}{{\ttfamily arXiv:1012.3210 [hep-th]}}.

\bibitem{Hama:2010av}
N.~Hama, K.~Hosomichi, and S.~Lee, ``{Notes on SUSY Gauge Theories on
  Three-Sphere},'' \href{http://dx.doi.org/10.1007/JHEP03(2011)127}{{\em JHEP}
  {\bfseries 03} (2011) 127},
\href{http://arxiv.org/abs/1012.3512}{{\ttfamily arXiv:1012.3512 [hep-th]}}.

\bibitem{Jafferis:2011zi}
D.~L. Jafferis, I.~R. Klebanov, S.~S. Pufu, and B.~R. Safdi, ``{Towards the
  F-Theorem: N=2 Field Theories on the Three-Sphere},''
  \href{http://dx.doi.org/10.1007/JHEP06(2011)102}{{\em JHEP} {\bfseries 06}
  (2011) 102},
\href{http://arxiv.org/abs/1103.1181}{{\ttfamily arXiv:1103.1181 [hep-th]}}.

\bibitem{Closset:2012vg}
C.~Closset, T.~T. Dumitrescu, G.~Festuccia, Z.~Komargodski, and N.~Seiberg,
  ``{Contact Terms, Unitarity, and F-Maximization in Three-Dimensional
  Superconformal Theories},''
  \href{http://dx.doi.org/10.1007/JHEP10(2012)053}{{\em JHEP} {\bfseries 10}
  (2012) 053},
\href{http://arxiv.org/abs/1205.4142}{{\ttfamily arXiv:1205.4142 [hep-th]}}.

\bibitem{Martelli:2005tp}
D.~Martelli, J.~Sparks, and S.-T. Yau, ``{The Geometric dual of a-maximisation
  for Toric Sasaki-Einstein manifolds},''
  \href{http://dx.doi.org/10.1007/s00220-006-0087-0}{{\em Commun. Math. Phys.}
  {\bfseries 268} (2006) 39--65},
\href{http://arxiv.org/abs/hep-th/0503183}{{\ttfamily arXiv:hep-th/0503183
  [hep-th]}}.

\bibitem{Amariti:2012tj}
A.~Amariti and S.~Franco, ``{Free Energy vs Sasaki-Einstein Volume for Infinite
  Families of M2-Brane Theories},''
  \href{http://dx.doi.org/10.1007/JHEP09(2012)034}{{\em JHEP} {\bfseries 09}
  (2012) 034},
\href{http://arxiv.org/abs/1204.6040}{{\ttfamily arXiv:1204.6040 [hep-th]}}.

\bibitem{Lee:2014rca}
S.~Lee and D.~Yokoyama, ``{Geometric free energy of toric AdS$_{4}$/CFT$_{3}$
  models},'' \href{http://dx.doi.org/10.1007/JHEP03(2015)103}{{\em JHEP}
  {\bfseries 03} (2015) 103},
\href{http://arxiv.org/abs/1412.8703}{{\ttfamily arXiv:1412.8703 [hep-th]}}.

\bibitem{Cacciatori:2009iz}
S.~L. Cacciatori and D.~Klemm, ``{Supersymmetric AdS(4) black holes and
  attractors},'' \href{http://dx.doi.org/10.1007/JHEP01(2010)085}{{\em JHEP}
  {\bfseries 01} (2010) 085},
\href{http://arxiv.org/abs/0911.4926}{{\ttfamily arXiv:0911.4926 [hep-th]}}.

\bibitem{Andrianopoli:1996cm}
L.~Andrianopoli, M.~Bertolini, A.~Ceresole, R.~D'Auria, S.~Ferrara, P.~Fre, and
  T.~Magri, ``{N=2 supergravity and N=2 superYang-Mills theory on general
  scalar manifolds: Symplectic covariance, gaugings and the momentum map},''
  \href{http://dx.doi.org/10.1016/S0393-0440(97)00002-8}{{\em J. Geom. Phys.}
  {\bfseries 23} (1997) 111--189},
\href{http://arxiv.org/abs/hep-th/9605032}{{\ttfamily arXiv:hep-th/9605032
  [hep-th]}}.

\bibitem{Ferrara:1996dd}
S.~Ferrara and R.~Kallosh, ``{Supersymmetry and attractors},''
  \href{http://dx.doi.org/10.1103/PhysRevD.54.1514}{{\em Phys. Rev.} {\bfseries
  D54} (1996) 1514--1524},
\href{http://arxiv.org/abs/hep-th/9602136}{{\ttfamily arXiv:hep-th/9602136
  [hep-th]}}.

\bibitem{Benini:2015eyy}
F.~Benini, K.~Hristov, and A.~Zaffaroni, ``{Black hole microstates in AdS$_{4}$
  from supersymmetric localization},''
  \href{http://dx.doi.org/10.1007/JHEP05(2016)054}{{\em JHEP} {\bfseries 05}
  (2016) 054},
\href{http://arxiv.org/abs/1511.04085}{{\ttfamily arXiv:1511.04085 [hep-th]}}.

\bibitem{Benini:2016rke}
F.~Benini, K.~Hristov, and A.~Zaffaroni, ``{Exact microstate counting for
  dyonic black holes in AdS4},''
  \href{http://dx.doi.org/10.1016/j.physletb.2017.05.076}{{\em Phys. Lett.}
  {\bfseries B771} (2017) 462--466},
\href{http://arxiv.org/abs/1608.07294}{{\ttfamily arXiv:1608.07294 [hep-th]}}.

\bibitem{Cvetic:2005zi}
M.~Cvetic, G.~W. Gibbons, H.~Lu, and C.~N. Pope, ``{Rotating black holes in
  gauged supergravities: Thermodynamics, supersymmetric limits, topological
  solitons and time machines},''
\href{http://arxiv.org/abs/hep-th/0504080}{{\ttfamily arXiv:hep-th/0504080
  [hep-th]}}.

\bibitem{Hristov:2019mqp}
K.~Hristov, S.~Katmadas, and C.~Toldo, ``{Matter-coupled supersymmetric
  Kerr-Newman-AdS$_4$ black holes},''
  \href{http://dx.doi.org/10.1103/PhysRevD.100.066016}{{\em Phys. Rev.}
  {\bfseries D100} no.~6, (2019) 066016},
\href{http://arxiv.org/abs/1907.05192}{{\ttfamily arXiv:1907.05192 [hep-th]}}.

\bibitem{DallAgata:2010ejj}
G.~Dall'Agata and A.~Gnecchi, ``{Flow equations and attractors for black holes
  in N = 2 U(1) gauged supergravity},''
  \href{http://dx.doi.org/10.1007/JHEP03(2011)037}{{\em JHEP} {\bfseries 03}
  (2011) 037},
\href{http://arxiv.org/abs/1012.3756}{{\ttfamily arXiv:1012.3756 [hep-th]}}.

\bibitem{Benini:2015noa}
F.~Benini and A.~Zaffaroni, ``{A topologically twisted index for
  three-dimensional supersymmetric theories},''
  \href{http://dx.doi.org/10.1007/JHEP07(2015)127}{{\em JHEP} {\bfseries 07}
  (2015) 127},
\href{http://arxiv.org/abs/1504.03698}{{\ttfamily arXiv:1504.03698 [hep-th]}}.

\bibitem{Jeffrey:1993}
L.~Jeffrey and F.~Kirwan, ``{Localization for nonabelian group actions},'' {\em
  Topology} {\bfseries 34} (1995) 291--327,
  \href{http://arxiv.org/abs/alg-geom/9307001}{{\ttfamily
  arXiv:alg-geom/9307001 [alg-geom]}}.

\bibitem{Benini:2013nda}
F.~Benini, R.~Eager, K.~Hori, and Y.~Tachikawa, ``{Elliptic genera of
  two-dimensional N=2 gauge theories with rank-one gauge groups},''
  \href{http://dx.doi.org/10.1007/s11005-013-0673-y}{{\em Lett. Math. Phys.}
  {\bfseries 104} (2014) 465--493},
\href{http://arxiv.org/abs/1305.0533}{{\ttfamily arXiv:1305.0533 [hep-th]}}.

\bibitem{Benini:2013xpa}
F.~Benini, R.~Eager, K.~Hori, and Y.~Tachikawa, ``{Elliptic Genera of 2d
  ${\mathcal{N}}$ = 2 Gauge Theories},''
  \href{http://dx.doi.org/10.1007/s00220-014-2210-y}{{\em Commun. Math. Phys.}
  {\bfseries 333} no.~3, (2015) 1241--1286},
\href{http://arxiv.org/abs/1308.4896}{{\ttfamily arXiv:1308.4896 [hep-th]}}.

\bibitem{Hori:2014tda}
K.~Hori, H.~Kim, and P.~Yi, ``{Witten Index and Wall Crossing},''
  \href{http://dx.doi.org/10.1007/JHEP01(2015)124}{{\em JHEP} {\bfseries 01}
  (2015) 124},
\href{http://arxiv.org/abs/1407.2567}{{\ttfamily arXiv:1407.2567 [hep-th]}}.

\end{thebibliography}\endgroup

\end{document}